\documentclass[amsmath,amssymb,aps,prx,reprint,superscriptaddress]{revtex4-2}

\usepackage{amsmath}
\usepackage{xcolor}
\usepackage{amsfonts,amssymb}
\usepackage{graphicx}
\usepackage{hyperref}
\usepackage{academicons}
\usepackage{multirow}
\usepackage{xr}

\begin{document}
\externaldocument[supp-]{supplementary_information}
\title{CHGNet: Pretrained universal neural network potential for charge-informed atomistic modeling}



\author{\text{Bowen Deng}}
\affiliation{Department of Materials Science and Engineering, University of California, Berkeley, California 94720, United States}
\affiliation{Materials Sciences Division, Lawrence Berkeley National Laboratory, California 94720, United States}

\author{\text{Peichen Zhong}}
\email[]{zhongpc@berkeley.edu}
\affiliation{Department of Materials Science and Engineering, University of California, Berkeley, California 94720, United States}
\affiliation{Materials Sciences Division, Lawrence Berkeley National Laboratory, California 94720, United States}

\author{\text{KyuJung Jun}}
\affiliation{Department of Materials Science and Engineering, University of California, Berkeley, California 94720, United States}
\affiliation{Materials Sciences Division, Lawrence Berkeley National Laboratory, California 94720, United States}

\author{\text{Janosh Riebesell}}
\affiliation{Materials Sciences Division, Lawrence Berkeley National Laboratory, California 94720, United States}
\affiliation{Cavendish Laboratory, University of Cambridge, United Kingdom}

\author{\text{Kevin Han}}
\affiliation{Materials Sciences Division, Lawrence Berkeley National Laboratory, California 94720, United States}

\author{\text{Christopher J. Bartel}}
\affiliation{Department of Materials Science and Engineering, University of California, Berkeley, California 94720, United States}
\affiliation{Department of Chemical Engineering and Materials Science, University of Minnesota, Minneapolis, Minnesota 55455, United States}

\author{\text{Gerbrand Ceder}}
\email[]{gceder@berkeley.edu}
\affiliation{Department of Materials Science and Engineering, University of California, Berkeley, California 94720, United States}
\affiliation{Materials Sciences Division, Lawrence Berkeley National Laboratory, California 94720, United States}

\date{\today}

\begin{abstract}
The simulation of large-scale systems with complex electron interactions remains one of the greatest challenges for the atomistic modeling of materials. Although classical force fields often fail to describe the coupling between electronic states and ionic rearrangements, the more accurate \textit{ab-initio} molecular dynamics suffers from computational complexity that prevents long-time and large-scale simulations, which are essential to study many technologically relevant phenomena, such as reactions, ion migrations, phase transformations, and degradation. 

In this work, we present the Crystal Hamiltonian Graph neural Network (CHGNet) as a novel machine-learning interatomic potential (MLIP), using a graph-neural-network-based force field to model a universal potential energy surface. CHGNet is pretrained on the energies, forces, stresses, and magnetic moments from the Materials Project Trajectory Dataset, which consists of over 10 years of density functional theory static and relaxation trajectories of $\sim 1.5$ million inorganic structures. The explicit inclusion of magnetic moments enables CHGNet to learn and accurately represent the orbital occupancy of electrons, enhancing its capability to describe both atomic and electronic degrees of freedom. We demonstrate several applications of CHGNet in solid-state materials, including charge-informed molecular dynamics in Li$_x$MnO$_2$, the finite temperature phase diagram for Li$_x$FePO$_4$ and Li diffusion in garnet conductors. We critically analyze the significance of including charge information for capturing appropriate chemistry, and we provide new insights into ionic systems with additional electronic degrees of freedom that can not be observed by previous MLIPs.

\end{abstract}


\maketitle

\section{Introduction}

\begin{figure*}[t]
\centering
\includegraphics[width=\linewidth]{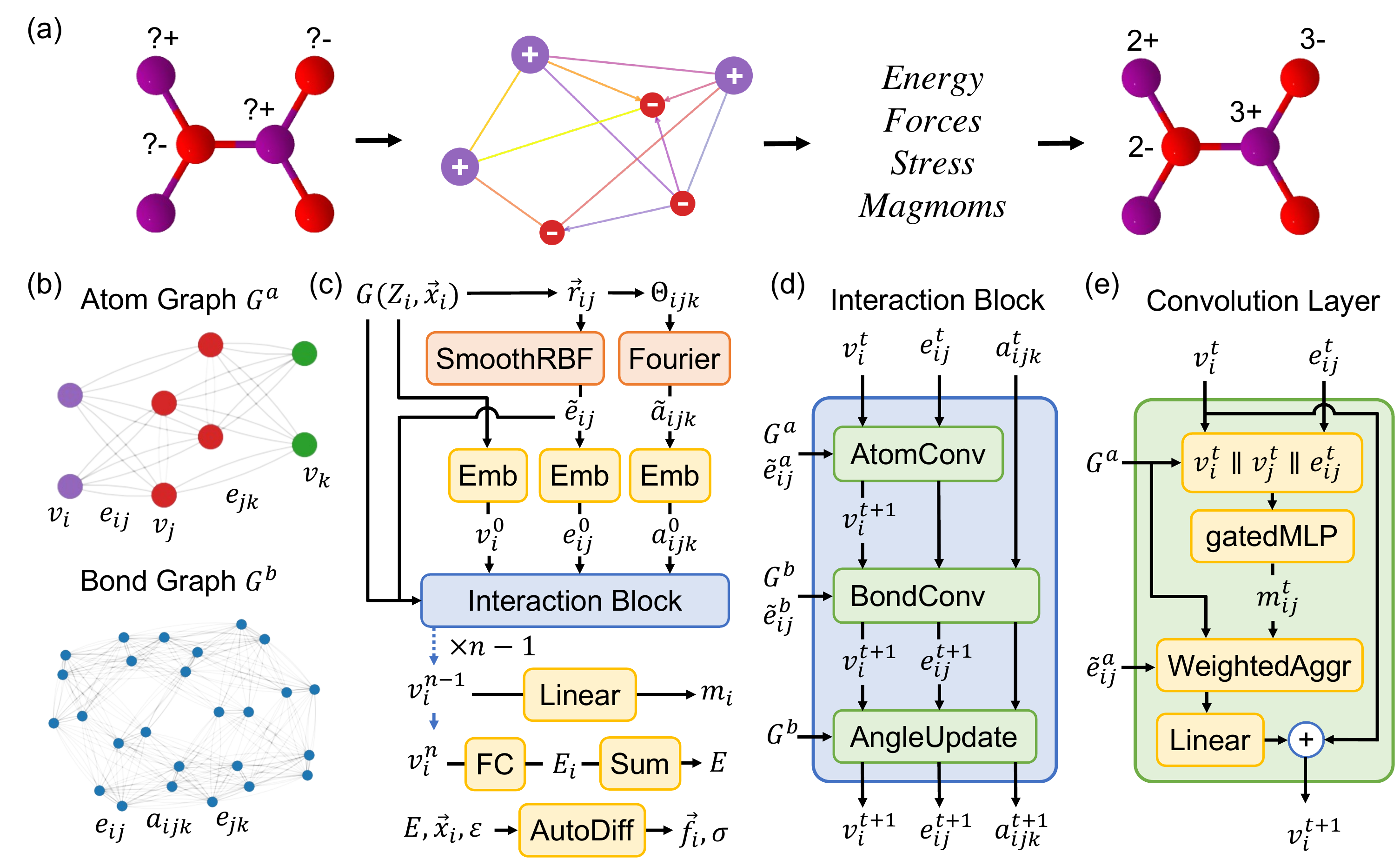}
\caption{\textbf{CHGNet model architecture} (a) CHGNet workflow: a crystal structure with unknown atomic charge is used as input to predict the energy, force, stress, and magnetic moments, resulting in a charge-decorated structure. (b) Atom graph: The pairwise bond information is drawn between atoms;  Bond graph: the pairwise angle information is drawn between bonds. (c) Graphs run through basis expansions and embedding layers to create atom, bond, angle features. The features are updated through several interaction blocks, and the properties are predicted at output layers. (d) Interaction block in which the atom, bond, and angle share and update information. (e) Atom convolution layer where neighboring atom and bond information is calculated through weighted message passing and aggregates to the atoms.}
\label{fig:CHGNet} 
\end{figure*}

Large-scale simulations, such as molecular dynamics (MD), are essential tools in the computational exploration of solid-state materials \cite{frenkel2001_understandMD}. They enable the study of reactivity, degradation, interfacial reactions, transport in partially disordered structures, and other heterogeneous phenomena relevant for the application of complex materials in technology. Technological relevance of such simulations requires rigorous chemical specificity which originates from the orbital occupancy of atoms. Despite their importance, accurate modeling of electron interactions or their subtle effects in MD simulations remains a major challenge. Classical force fields treat the charge as an atomic property that is assigned to every atom \textit{a-priori} \cite{Lucas_Bauer_Patel_2012, Drautz2020_ACE}. Methodology developments in the field of polarizable force fields such as the electronegativity equalization method (EEM) \cite{Mortier_Ghosh_Shankar_1986}, chemical potential equalization (CPE) \cite{York_Yang_1996}, and charge equilibration (Qeq) \cite{Rappe_Goddard_1991} realize charge evolution via the redistribution of atomic partial charge. However, these empirical methods are often not accurate enough to capture complex electron interactions.

\textit{Ab-initio} molecular dynamics (AIMD) with density functional theory (DFT) can produce high-fidelity results with quantum-mechanical accuracy by explicitly computing the electronic structure within the density functional approximation. The charge-density distribution and corresponding energy can be obtained by solving the Kohn--Sham equation \cite{KohnSham}. Long-time and large-scale spin-polarized AIMD simulations critical for studying ion migrations, phase transformations and chemical reactions are challenging and extremely computing intensive \cite{Reed2004_ChemicalReview, Eum2020_TM_migration_layer}. These difficulties underscore the need for more efficient computational methods in the field that can account for charged ions and their orbital occupancy at sufficient time and length scales needed to model important phenomena.

Machine-learning interatomic potentials (MLIPs) such as \ae net \cite{Artrith_Morawietz_Behler_2011, Nong_aenet_GPU_2023} and DeepMD \cite{Zhang2021_DeepMD_water} have provided promising solutions to bridge the gap between expensive electronic structure methods and efficient classical interatomic potentials. Specifically, graph neural network (GNN)-based MLIPs such as DimeNet \cite{dimnet_2020}, NequIP \cite{NequIP_2022}, TeaNet \cite{TeaNet_2022}, and MACE \cite{Batatia2022_MACE} have been shown to achieve state-of-the-art performance by incorporating invariant/equivariant symmetry constraints and long-range interaction through graph convolution \cite{Fu2022_forces}.
Most recently, GNN-based MLIPs trained on the periodic table (e.g., M3GNet) have demonstrated the possibility of universal interatomic potentials that may not require chemistry-specific training for each new application \cite{Chen_Ong_2022, Choudhary2023_universal, PFP_2022}. However, the inclusion of the important effects that valences have on chemical bonding remains a challenge for MLIPs, and the early success derived mostly from the inclusion of electrostatics for long-range interactions \cite{spookynet, charge-nequip, AIMNet-NSE}. 

The importance of an ion's valence derives from the fact that it can engage in very different bonding with its environment depending on its electron count. While traditional MLIPs treat the elemental label as the basic chemical identity, different valence states of transition metal ions behave as different from each other as different elements. For example, high spin Mn$^{4+}$ is a non-bonding spherical ion which almost always resides in octahedral coordination by oxygen atoms, whereas Mn$^{3+}$ is a Jahn--Teller active ion that radically distorts its environment, and Mn$^{2+}$ is an ion that strongly prefers tetrahedral coordination \cite{Reed2004_ChemicalReview}. Such strong chemical interaction variability across different valence states exists for almost all transition metal ions and requires specification of an ion beyond its chemical identity. In addition, the charge state is a degree of freedom that can create configurational entropy and whose dynamic optimization can lead to strongly coupled charge and ion motion, which is impossible to capture with a MLIP that only carries elemental labels. The relevance of explicit electron physics motivates the development of a robust MLIP model with charge information built in.

Charge has been represented in a variety of ways, from a simple oxidation state label to continuous wave functions derived from quantum mechanics \cite{Walsh2018_chg_review}. Challenges in incorporating charge information into MLIPs arise from many factors, such as the ambiguity of representations \cite{Xie_Persson_Small_2020}, complexity of interpretation \cite{Gong2019_charge_density}, scarcity of labels \cite{charge-nequip}, and impracticality of taking charge as an input ($E(\{\boldsymbol{r}_i\}, \{q_i\})$, as the labels $\{q_i\}$ are generally not \textit{a-priori} available) \cite{spookynet}. In this work, we define charge as an atomic property (\emph{atomic charge}) that can be inferred from the inclusion of magnetic moments (magmoms). We show that by explicitly incorporating the site-specific magmoms as the charge-state constraints into the \textbf{C}rystal \textbf{H}amiltonian \textbf{G}raph neural-\textbf{Net}work (CHGNet), one can both enhance the latent-space regularization and accurately capture electron interactions. 

We demonstrate the charge constraints and latent-space regularization of atomic charge in Na$_2$V$_2$(PO$_4$)$_3$ and show the applications of CHGNet in the study of charge-transfer and phase transformation in Li$_x$MnO$_2$, electronic entropy in the Li$_x$FePO$_4$ phase diagram, and Li diffusivity in garnet-type Li-superionic conductors Li$_{3+x}$La$_3$Te$_2$O$_{12}$. By critically comparing and evaluating the importance of incorporating charge information in the construction of CHGNet, we offer new insights into the materials modeling of ionic systems with additional electronic degrees of freedom. Our analysis highlights the essential role that charge information plays in atomistic simulations for solid-state materials.

\section{Results}
\begin{figure*}[t]
\centering
\includegraphics[width=\linewidth]{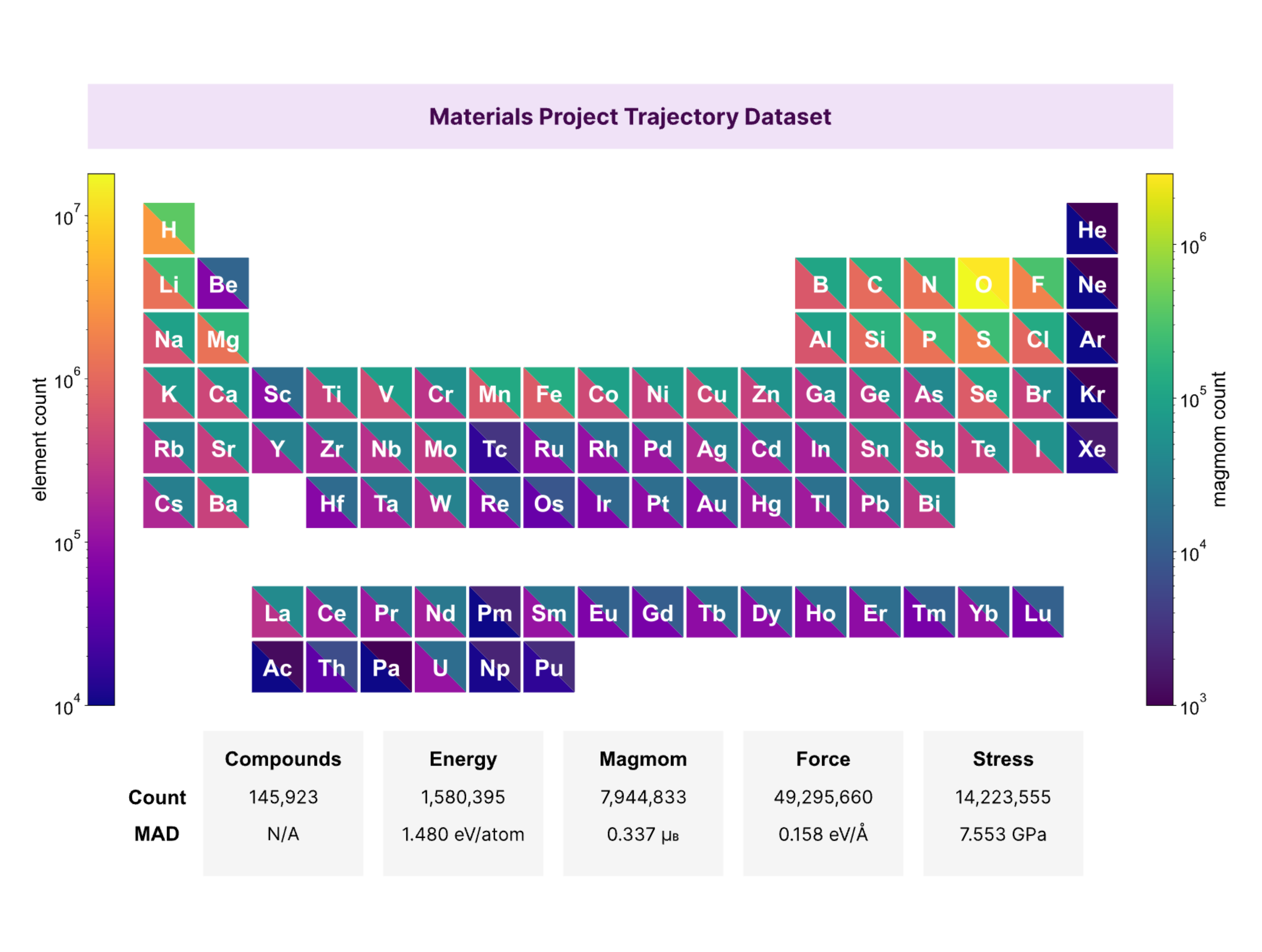}
\caption{\textbf{Element distribution of Materials Project Trajectory (MPtrj) Dataset.} The color on the lower-left triangle indicates the total number of atoms/ions of an element. The color on the upper right indicates the number of times the atoms/ions are incorporated with magnetic moment labels in the MPtrj dataset. On the lower part of the plot is the count and mean absolute deviation (MAD) of energy, magmoms, force, and stress}
\label{fig:MPtrj} 
\end{figure*}

\subsection{CHGNet architecture}
The foundation of CHGNet is a GNN, as shown in Fig. \ref{fig:CHGNet}, where the graph convolution layer is used to propagate atomic information via a set of nodes $\{v_i\}$ connected by edges $\{e_{ij}\}$. The translation, rotation, and permutation invariance are preserved in GNNs \cite{Bruna_Zaremba_Szlam_LeCun_2013, Geiger_Smidt_2022, Xie_Grossman_2018}. 
Figure \ref{fig:CHGNet}(a) shows the workflow of CHGNet which takes a crystal structure with unknown atomic charges as input and outputs the corresponding energy, forces, stress, and magmoms. The charge-decorated structure can be inferred from the on-site magnetic moments and atomic orbital theory. The details are described in the following section.

In CHGNet, a periodic crystal structure is converted into an atom graph $G^a$ by searching for neighboring atoms $v_j$ within $r_\text{cut}$ of each atom $v_i$ in the primitive cell. The edges $e_{ij}$ are drawn with information from the pairwise distance between $v_i$ and $v_j$, as shown in Fig. \ref{fig:CHGNet}(b). Three-body interaction can be computed by using an auxiliary bond graph $G^b$, which can be similarly constructed by taking the angle $a_{ijk}$ as the pairwise information between bonds $e_{ij}$ and $e_{jk}$ (see \hyperref[sec:methods]{Methods}). We adopt similar approaches to include the angular/three-body information as other recent GNN MLIPs \cite{dimnet_2020, Chen_Ong_2022, Choudhary_DeCost_2021}. 

Figure \ref{fig:CHGNet}(c) shows the architecture of CHGNet, which consists of a sequence of basis expansions, embeddings, interaction blocks, and outputs layers (see Methods for details). 
Figure \ref{fig:CHGNet}(d) illustrates the components within an interaction block, where the atomic interaction is simulated with the update of atom, bond, and angle features via the convolution layers. 
Figure \ref{fig:CHGNet}(e) presents the convolution layer in the atom graph. Weighted message passing is used to propagate information between atoms, where the message weight $\Tilde{e}_{ij}^a$ from node $j$ to node $i$ decays to zero at the graph cutoff radius to ensure smoothness of the potential energy surface \cite{dimnet_2020}.

Unlike other GNNs, where the updated atom features $\{v^n_i\}$ after $n$ convolution layers are directly used to predict energies, CHGNet regularizes the node-wise features $\{v^{n-1}_i\}$ at the $n-1$ convolution layer to contain the information about magnetic moments. The regularized features $\{v^{n-1}_i\}$ carry rich information about both local ionic environments and charge distribution. Therefore, the atom features $\{v^n_i\}$ used to predict energy, force, and stress are charge-constrained by their charge-state information. As a result, CHGNet can provide charge-state information using only the nuclear positions and atomic identities as input, allowing the study of charge distribution in atomistic modeling.

\subsection{Materials Project Trajectory Dataset}

The Materials Project database contains a vast collection of DFT calculations on $\sim 146,000$ inorganic materials composed of 94 elements \cite{MaterialProject}. To accurately sample the universal potential energy surface, we extracted $\sim 1.37$ million Materials Project tasks of structure relaxation and static calculations using either the generalized gradient approximation (GGA) or GGA$+U$ exchange-correlation (see \hyperref[sec:methods]{Methods}). This effort resulted in a comprehensive dataset with 1,580,395 atom configurations, 1,580,395 energies, 7,944,833 magnetic moments, 49,295,660 forces, and 14,223,555 stresses. To ensure the consistency of energies within the MPtrj dataset, we applied the GGA/GGA$+U$ mixing compatibility correction, as described by \citet{Wang_Kingsbury_McDermott_Horton_Jain_Ong_Dwaraknath_Persson_2021}.

The distribution of elements in the MPtrj dataset is illustrated in Fig. \ref{fig:MPtrj}. The lower-left triangle (warm color) in an element's box indicates the frequency of occurrence of that element in the dataset, and the upper-right triangle (cold color) represents the number of instances where magnetic information is available for the element. With over 100,000 occurrences for 60 different elements and more than 10,000 instances with magnetic information for 76 different elements, the MPtrj dataset provides comprehensive coverage of all chemistries, excluding only the noble gases and actinoids. The lower boxes in Fig. \ref{fig:MPtrj} present the counts and mean absolute deviations of energy, force, stress, and magmoms in the MPtrj dataset.

\subsection{Performance Evaluation}

CHGNet with 400,438 trainable parameters was trained on the MPtrj dataset with an 8:1:1 training, validation, and test set ratio, partitioned by materials (see \hyperref[sec:methods]{Methods}). The statistics of the mean absolute errors of the energy, force, stress, and magmoms on the MPtrj test set structures are shown in Table \ref{table:1}. We observe a similar test set error with slight improvements in the model trained with magmoms.

To evaluate the robustness of CHGNet as a universal force field, we submitted CHGNet to Matbench Discovery \footnote{J. Riebesell, Rhy. Goodall, A. Jain, K. Persson, A. Lee, Matbench Discovery: Can machine learning identify stable crystals? \href{https://matbench-discovery.materialsproject.org/}{https://matbench-discovery.materialsproject.org}} for out-of-distribution material stability prediction. CHGNet achieves state-of-the-art performance in high-throughput stable inorganic crystal discovery compared to 8 other models submitted to this benchmark at the time of writing.

For a benchmark on CHGNet molecular dynamic simulations, we applied the pretrained CHGNet to MD simulations on Li superionic conductors that were previously reported with AIMD \cite{Jun_Sun_Xiao_Zeng_Kim_Kim_Miara_Im_Wang_Ceder_2022}. Supplementary Fig. 1 shows that CHGNet systematically agrees with AIMD results on room temperature conductivities and activation energies within the AIMD error bar, and CHGNet successfully distinguishes the faster and slower conductors that were identified with DFT.

\begin{table}[h]
\caption{Mean-absolute-errors (MAEs) of CHGNet on MPtrj test set of 157,955 structures from 14,572 materials. 'With mag' and 'No mag' indicate whether the model is trained with magmoms ($\mu_B$ is the Bohr magneton).}
\label{sample-table}
\begin{center}
\begin{tabular}{clllll}
\hline\hline
\multicolumn{1}{c}{}  & 
\multicolumn{1}{c}{\begin{tabular}{@{}c@{}} $\textbf{Energy}$ \\ (meV/atom) \end{tabular}} & 
\multicolumn{1}{c}{\begin{tabular}{@{}c@{}} $\textbf{Force}$ \\ (meV/\AA) \end{tabular}} &
\multicolumn{1}{c}{\begin{tabular}{@{}c@{}} $\textbf{Stress}$ \\ (GPa) \end{tabular}} &
\multicolumn{1}{c}{\begin{tabular}{@{}c@{}} $\textbf{Magmom}$ \\ ($\mu_B$) \end{tabular}} 
\\ \hline 
\textbf{With mag}   & \multicolumn{1}{c}{30} & \multicolumn{1}{c}{77} & \multicolumn{1}{c}{0.348} & \multicolumn{1}{c}{0.032} \\
\hline 
\textbf{No mag}   & \multicolumn{1}{c}{33} & \multicolumn{1}{c}{79} & \multicolumn{1}{c}{0.351} & \multicolumn{1}{c}{ N/A } \\
\hline\hline
\end{tabular}
\end{center}
\label{table:1}
\end{table}

\subsection{Charge-constraints and charge-inference from magnetic moments}

\begin{figure}[htb]
\centering
\includegraphics[width=\linewidth]{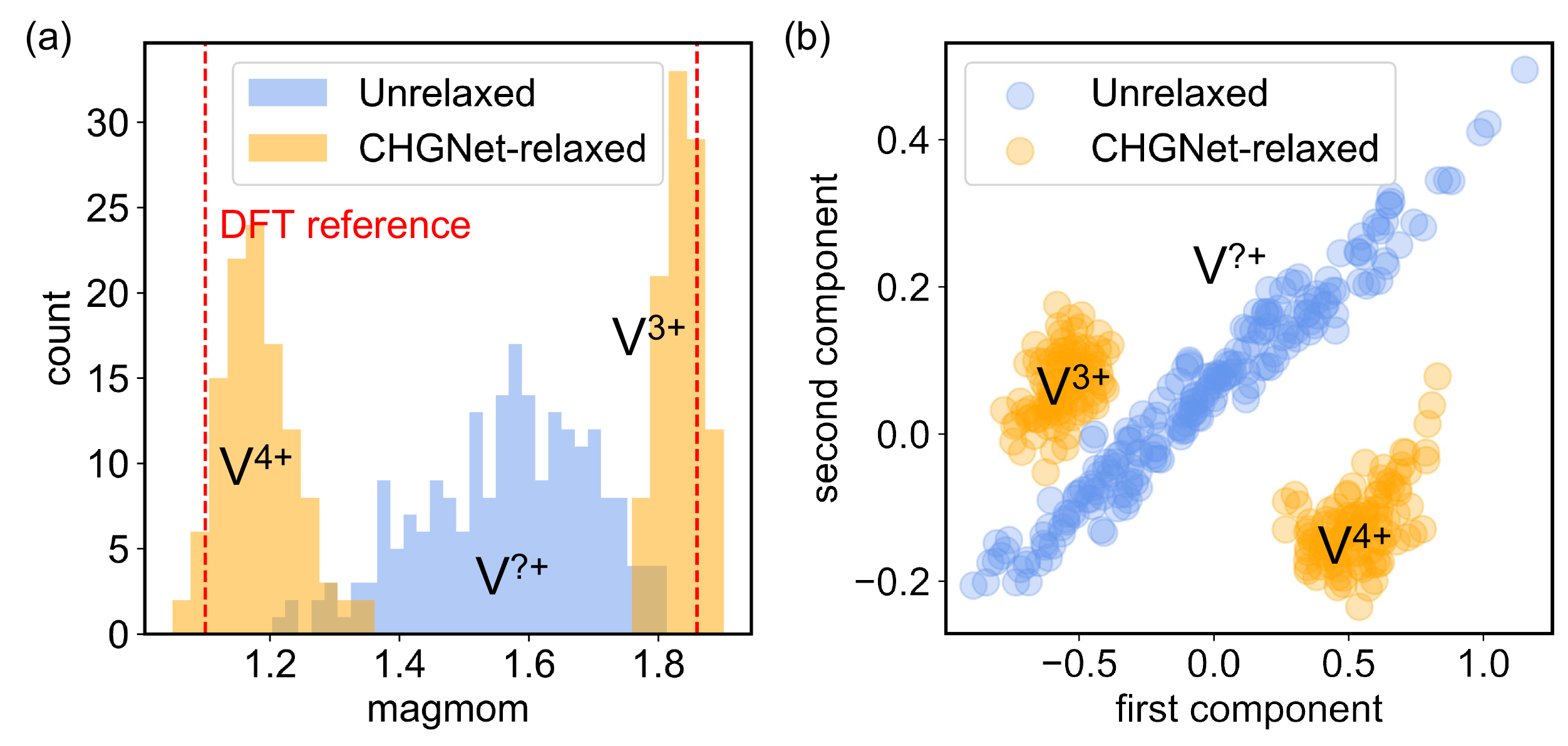}
\caption{\textbf{Magmom and hidden-space regularization in Na$_2$V$_2$(PO$_4$)$_3$}. (a) Magmom distribution of the 216 V ions in the unrelaxed structure (blue) and CHGNet-relaxed structure (orange). (b) A two-dimensional visualization of the PCA on V-ion embedding vectors before the magmom projection layer indicates the latent space clustering is highly correlated with magmoms and charge information. The PCA reduction is calculated for both unrelaxed and relaxed structures.} 
\label{fig:NVP} 
\end{figure}

\begin{figure*}[t]
\centering
\includegraphics[width=\linewidth]{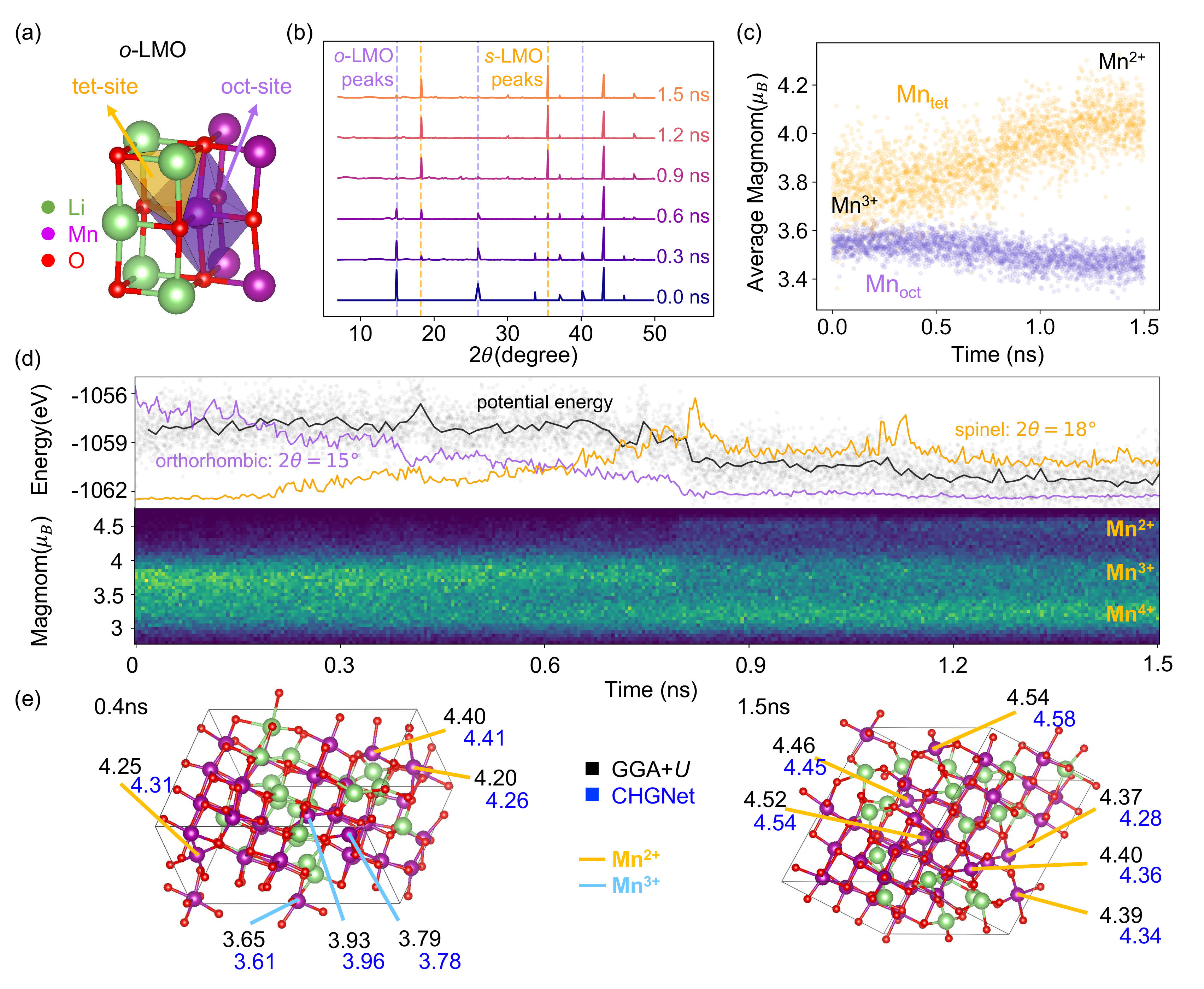}
\caption{\textbf{Li$_{0.5}$MnO$_2$ phase transformation and charge disproportionation} (a) orthorhombic LiMnO$_2$ (\textit{o}-LMO) unit cell plotted with the tetrahedral site and the octahedral site. (b) Simulated XRD pattern of CHGNet MD structures as the system transforms from the \textit{o}-LMO phase to the \textit{s}-LMO. (c) Average magmoms of tetrahedral and octahedral Mn ions \textit{vs.} time. (d) Top: total potential energy and the relative intensity of \textit{o}-LMO and \textit{s}-LMO characteristic peaks \textit{vs.} time. The solid black line is averaged over every 10 ps. Bottom: the histogram of magmoms on all Mn ions \textit{vs.} time. The brighter color indicates more Mn ions distributed at the magmom. (e) Predicted magmoms of tetrahedral Mn ions using GGA$+U$-DFT (black) and CHGNet (blue), where the structures are drawn from MD simulation at 0.4 ns (left) and 1.5 ns (right).}
\label{fig:LMO} 
\end{figure*}

In solid-state materials that contain heterovalent ions, it is crucial to distinguish the atomic charge of the ions, as an element's interaction with its environment can depend strongly on its valence state. It is well known that the valence of heterovalent ions cannot be directly calculated through the DFT charge density because the charge density is almost invariant to the valence state due to the hybridization shift with neighboring ligand ions \cite{Mackrodt1996_HF_chg,Wolverton1998_LiCO2}. Furthermore, the accurate representation and encoding of the full charge density is another demanding task requiring substantial computational resources \cite{Gong2019_charge_density,Qiao2022_OrbitEqui}. An established approach is to rely on the magnetic moment for a given atom site as an indicator of its atomic charge, which can be derived from the difference in localized up-spin and down-spin electron densities in spin-polarized DFT calculations \cite{Reed2004_ChemicalReview,Barroso-Luque2022_CE_theory}. Compared with the direct use of charge density, magmoms are found to contain more comprehensive information regarding the electron orbital occupancy and therefore the chemical behavior of ions, as demonstrated in previous studies.

To rationalize our treatment of the atomic charge, we used a NASICON-type Na-ion cathode material Na$_4$V$_2$(PO$_4$)$_3$ as an illustrative example. The phase stability of the (de-)intercalated material Na$_{4-x}$V$_2$(PO$_4$)$_3$ is associated with Na/vacancy ordering and is highly correlated to the charge ordering on the vanadium sites \cite{Wang2022_NVP_phase_stability}. We generated a supercell structure of Na$_4$V$_2$(PO$_4$)$_3$ with 2268 atoms and randomly removed half of the Na ions to generate the structure with composition Na$_2$V$_2$(PO$_4$)$_3$, where half of the V ions are oxidized to a V$^{4+}$ state. We used CHGNet to relax the (de-)intercalated structure and analyze its capability to distinguish the valence states of V atoms with the ionic relaxation (see \hyperref[sec:methods]{Methods}).

Figure \ref{fig:NVP}(a) shows the distribution of predicted magmoms on all V ions in the unrelaxed (blue) and relaxed (orange) structures. Without any prior knowledge about the V-ion charge distribution other than learning from the spatial coordination of the V nuclei, CHGNet successfully differentiated the V ions into two groups of V$^{3+}$ and V$^{4+}$. Figure \ref{fig:NVP}(b) shows the two-dimensional principal component analysis (PCA) of all the latent space feature vectors of V ions for both unrelaxed and relaxed structures after three interaction blocks. The PCA analysis demonstrates two well-separated distributions, indicating the latent space feature vectors of V ions are strongly correlated to the different valence states of V. Hence, imposing different magmom labels to the latent space (i.e., forcing the two orange peaks to converge to the red dashed lines in Fig. \ref{fig:NVP}(a)) would act as the \emph{charge constraints} for the model by regularizing the latent-space features.

Because energy, force, and stress are calculated from the same feature vectors, the inclusion of magmoms can improve the featurization of the heterovalent atoms in different local chemical environments (e.g., V$^{3+}$ and V$^{4+}$ displays very distinct physics and chemistry) and therefore improve the accuracy and expressibility of CHGNet.

\subsection{Charge disproportionation in Li$_x$MnO$_2$ phase transformation}

The long-time and large-scale simulation of CHGNet enables studies of ionic rearrangements coupled with charge transfer \cite{Reed_Ceder_Ven_2001,Kang2006_Li_diffusion}, which is crucial for ion mobility and the accurate representation of the interaction between ionic species.
As an example, in the LiMnO$_2$ battery cathode material, transition-metal migration plays a central role in its phase transformations, which cause irreversible capacity loss \cite{Reimers_Fuller_Rossen_Dahn_1993, Koetschau_Richard_Dahn_Soupart_Rousche_1995}. The mechanism of Mn migration is strongly coupled with charge transfer, with Mn$^{4+}$ being an immobile ion, and Mn$^{3+}$ and  Mn$^{2+}$ generally considered to be more mobile \cite{Jang_Chou_Huang_Sadoway_Chiang_2003,Jo2019_layer_Mn_migration,radin2019_NE_Mn7}. The dynamics of the coupling of the electronic degrees of freedom with those of the ions has been challenging to study but is crucial to understand the phase transformation from orthorhombic LiMnO$_2$ (\textit{o}-LMO, shown in Fig. \ref{fig:LMO}(a)) to spinel LiMnO$_2$ (\textit{s}-LMO), as the time scale and computational cost of such phenomena are far beyond any possible \textit{ab-initio} methods. 

In early quasi-static \textit{ab-initio} studies, \citet{Reed_Ceder_Ven_2001} rationalized the remarkable speed at which the phase transformation proceeds at room temperature using a charge disproportionation mechanism: $2\text{Mn}^{3+}_{\text{oct}} \rightarrow \text{Mn}^{2+}_{\text{tet}} + \text{Mn}^{4+}_{\text{oct}}$, where the subscript indicates location in the tetrahedral or octahedral site of a face-centered cubic oxygen packing, as shown in Fig. \ref{fig:LMO}(a). The hypothesis based on DFT calculations was that Mn$^{2+}$ had a lower energy barrier for migration between tetrahedral and octahedral sites and preferred to occupy the tetrahedral site. The ability therefore for Mn to dynamically change its valence would explain its remarkable room temperature mobility. However, \citet{Jang_Chou_Huang_Sadoway_Chiang_2003} showed in a later magnetic characterization experiment that the electrochemically transformed spinel LiMnO$_2$ has lower-spin (high-valence) Mn ions on the tetrahedral sites, which suggested the possibility that Mn with higher valence can be stable on tetrahedral sites during the phase transformation.

To demonstrate the ability of CHGNet to fully describe such a process, we used CHGNet to run a charge-informed MD simulation at 1100 K for 1.5 ns (see \hyperref[sec:methods]{Methods}). The MD simulation started from a partially delithiated supercell structure with  the \textit{o}-LMO structure (Li$_{20}$Mn$_{40}$O$_{80}$), which is characterized by peaks at 15\textdegree, 26\textdegree, and 40\textdegree\ in the X-ray diffraction (XRD) pattern (the bottom line in Fig. \ref{fig:LMO}(b)). As the simulation proceeded, a phase transformation from orthorhombic ordering to spinel-like ordering was observed. Figure \ref{fig:LMO}(b) presents the simulated XRD pattern of MD structures at different time intervals from 0 to 1.5 ns, with a clear increase in the characteristic spinel peaks (18\textdegree, 35\textdegree) and a decrease in the orthorhombic peak. The simulated results agree well with the experimental in-situ XRD results \cite{Reimers_Fuller_Rossen_Dahn_1993, Jang_Chou_Huang_Sadoway_Chiang_2003}.

Figure \ref{fig:LMO}(d) presents the CHGNet-predicted energy of the LMO supercell structure as a function of simulation time, together with the peak strength at $2\theta=15^\circ$ and $18^\circ$. An explicit correlation between the structural transformation and energy landscape is observed. The predicted average potential energy of the spinel phase is approximately 50 meV/oxygen lower than that of the starting \textit{o}-LMO, suggesting that the phase transformation to spinel is indeed, thermodynamically favored.

The advantage of CHGNet is shown in its ability to predict charge-coupled physics, as evidenced by the lower plot in Fig. \ref{fig:LMO}(d). A histogram of the magmoms of all the Mn ions in the structure is presented against time. In the early part of the simulation, the magmoms of Mn ions are mostly distributed between 3$\mu_B$ and 4$\mu_B$, which correspond to Mn$^{4+}$ and Mn$^{3+}$.
At approximately 0.8 ns, there is a significant increase in the amount of Mn$^{2+}$, which is accompanied by a decrease in the potential energy and changes in the XRD peaks. Following this major transformation point, the Mn$^{3+}$ ions undergo charge disproportionation, resulting in the coexistence of Mn$^{2+}$, Mn$^{3+}$, and Mn$^{4+}$ in the transformed spinel-like structure.

One important observation from the long-time charge-informed MD simulation is the correlation between ionic rearrangements and the charge-state evolution. Specifically, we noticed that the time scale of charge disproportionation ($\sim$ ns for the emergence of Mn$^{2+}$) is far longer than the time scale of ion hops ($\sim$ ps for the emergence of Mn$_{\text{tet}}$), indicating that the migration of Mn to the tetrahedral coordination is less likely related to the emergence of Mn$^{2+}$. Instead, our result indicates that the emergence of Mn$^{2+}_{\text{tet}}$ is correlated to the formation of the long-range spinel-like ordering. Figure \ref{fig:LMO}(c) shows the average magmoms of Mn$_{\text{tet}}$ and Mn$_{\text{oct}}$ as a function of time. The result reveals that Mn$^{2+}_{\text{tet}}$ only forms over a long time period, which cannot be observed using any conventional simulation techniques. 

To further validate this hypothesis and the accuracy of CHGNet prediction, we used GGA$+U$ and r$^2$SCAN-DFT (see Supplementary Information) static calculations to get the magmoms of the structures at 0.4 and 1.5 ns, where the GGA$+U$ results are shown in Fig. \ref{fig:LMO}(e). CHGNet (blue) shows highly accurate agreement with GGA$+U$ magmoms (black) and infers the same Mn$_\text{tet}$ valence states.

\subsection{Electronic entropy effect in the phase diagram of Li$_x$FePO$_4$}

\begin{figure}[htb]
\centering
\includegraphics[width=\linewidth]{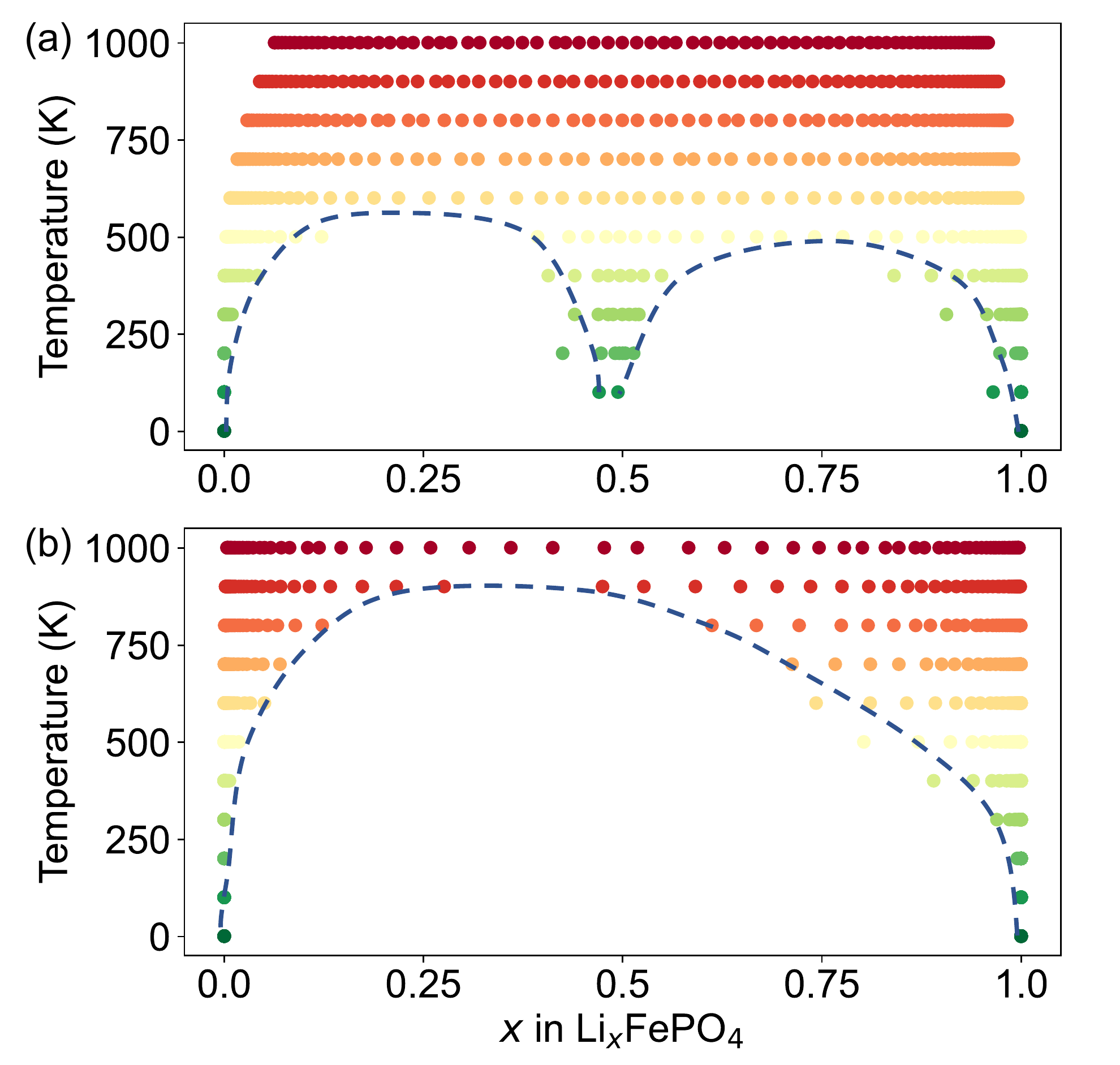}
\caption{\textbf{Li$_x$FePO$_4$ phase diagram from CHGNet}. The phase diagrams in (a) and (b) are calculated with and without electronic entropy on Fe$^{2+}$ and Fe$^{3+}$. The colored dots represent the stable phases obtained in semi-grand canonical MC. The dashed lines indicate the two-phase equilibria between solid solution phases.}
\label{fig:LFP} 
\end{figure}

The configurational electronic entropy has a significant effect on the temperature-dependent phase stability of mixed-valence oxides, and its equilibrium modeling therefore requires an explicit indication of the atomic charge. However, no current MLIPs can provide such information. We demonstrate that using CHGNet one can infer the atomic charge and include the electronic entropy in the computation of the temperature-dependent phase diagram (PD) of Li$_x$FePO$_4$.

Previous research has shown that the formation of a solid solution in Li$_x$FePO$_4$ is mainly driven by electronic entropy rather than by Li$^+$/vacancy configurational entropy \cite{Zhou_Maxisch_Ceder_2006}. We applied CHGNet as an energy calculator to generate two distinct cluster expansions (CEs), which is a typical approach to studying configurational entropy \cite{VandeWalle2002_ATAT}. One of these is charge-decorated (considering Li$^+$/vacancy and Fe$^{2+}$/Fe$^{3+}$) and another is non-charge-decorated (only 
considering Li$^+$/vacancy without consideration of the Fe valence). Semi-grand canonical Monte Carlo was used to sample these cluster expansions and construct Li$_x$FePO$_4$ PDs (see \hyperref[sec:methods]{Methods}). The calculated PD with charge decoration in Fig. \ref{fig:LFP}(a) features a miscibility gap between FePO$_4$ and LiFePO$_4$, with a eutectoid-like transition to the solid-solution phase at intermediate Li concentration, qualitatively matching the experiment result \cite{Delacourt_Poizot_Tarascon_Masquelier_2005, Dodd_Yazami_Fultz_2006}. In contrast, the calculated PD without charge decoration in Fig. \ref{fig:LFP}(b) features only a single miscibility gap without any eutectoid transitions, in disagreement with experiments. This comparison highlights the importance of explicit inclusion of the electronic degrees of freedom, as failure to do so can result in incorrect physics. These experiments show how practitioners may benefit from CHGNet with atomic charge inference for equilibrium modeling of configurationally and electronically disordered systems.

\subsection{Activated Li diffusion network in Li$_3$La$_3$Te$_2$O$_{12}$}
\begin{figure}[htb]
\centering
\includegraphics[width=\linewidth]{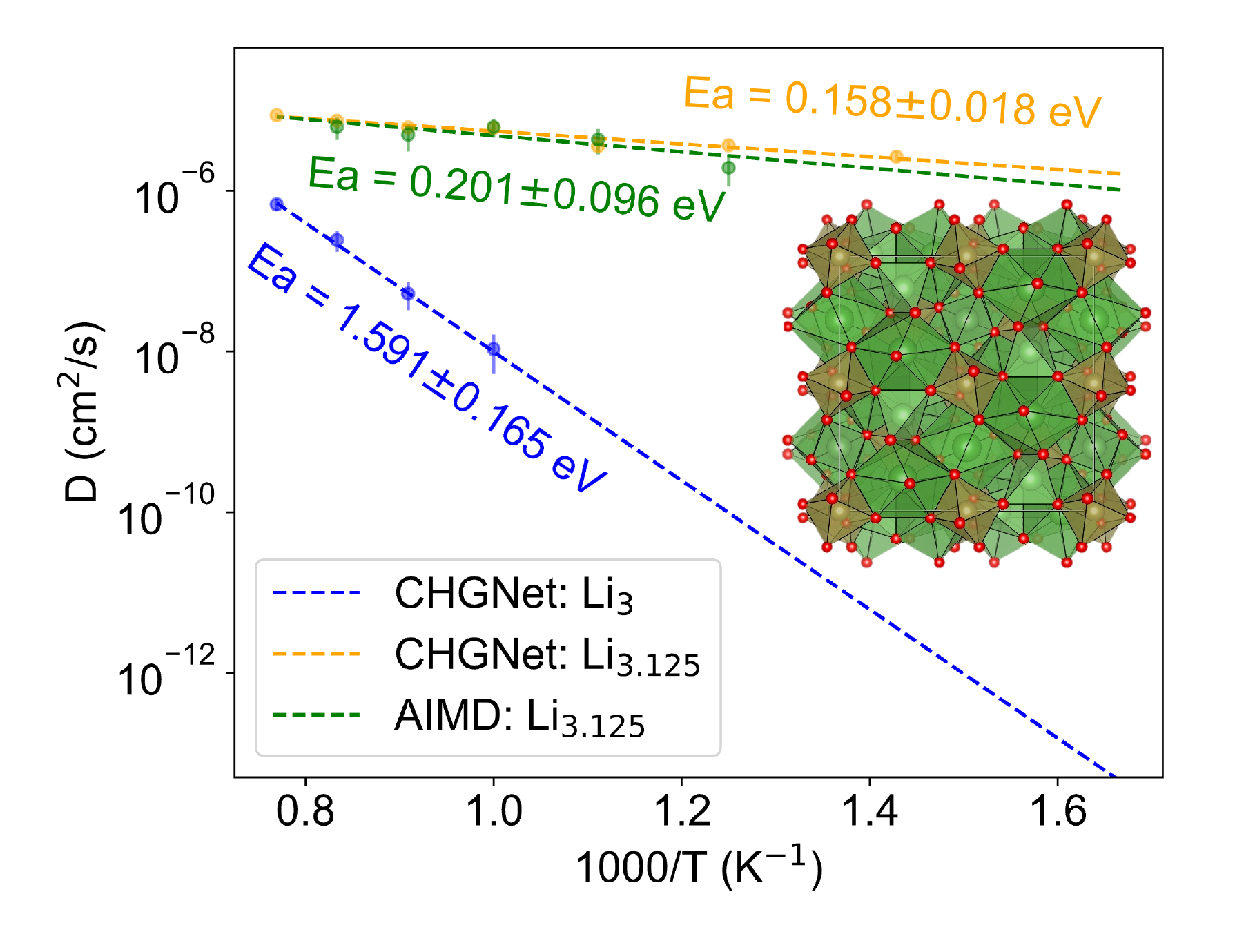}
\caption{\textbf{Li diffusivity in garnet Li$_3$La$_3$Te$_2$O$_{12}$}. The CHGNet simulation accurately reproduces the dramatic increase in Li-ion diffusivity when a small amount of extra Li is stuffed into the garnet structure, qualitatively matching the activated diffusion network theory and agreeing well with the DFT-computed activation energy.}
\label{fig:LLTO} 
\end{figure}

In this section, we showcase the precision of CHGNet for general-purpose MD. Lithium-ion diffusivity in fast Li-ion conductors is known to show a drastic non-linear response to compositional change. For example, stuffing a small amount of excess lithium into stoichiometric compositions can result in orders-of-magnitude improvement of the ionic conductivity \cite{Jun_Sun_Xiao_Zeng_Kim_Kim_Miara_Im_Wang_Ceder_2022}. \citet{Xiao_Jun_Wang_Miara_Tu_Ceder_2021} reported that the activation energy of Li diffusion in stoichiometric garnet Li$_3$La$_3$Te$_2$O$_{12}$ decreases from more than 1 eV to $\sim$160 meV in a slightly stuffed Li$_{3+\delta}$ garnet ($\delta$ = 1/48), owing to the activated Li diffusion network of face-sharing tetrahedral and octahedral sites. 

We performed a zero-shot test to assess the ability of CHGNet to capture the effect of such slight compositional change on the diffusivity and its activation energy. Figure \ref{fig:LLTO} shows the Arrhenius plot from CHGNet-based MD simulations and compares it to AIMD results. Our results indicate that not only is the activated diffusion network effect precisely captured, the activation energies from CHGNet are also in excellent agreement with the DFT results \cite{Xiao_Jun_Wang_Miara_Tu_Ceder_2021}. This effort demonstrates the capability of CHGNet to precisely capture the strong interactions between Li ions in activated local environments and the ability to simulate highly non-linear diffusion behavior. Moreover, CHGNet can dramatically decrease the error on simulated diffusivity and enable studies in systems with poor diffusivity such as the unstuffed Li$_{3}$ garnet by extending to nano-second-scale simulations \cite{Huang2021_DP}.

\section{Discussion}
Large-scale computational modeling has proven essential in providing atomic-level information in materials science, medical science, and molecular biology. Many technologically relevant applications contain heterovalent species, for which a comprehensive understanding of the atomic charge involved in the dynamics of processes is of great interest. 
The importance of assigning a valence to ions derives from the fundamentally different electronic and bonding behavior ions can exhibit when their electron count changes. \textit{Ab-initio} calculations based on DFT are useful for these problems, but the $\sim \mathcal{O}(N^3)$ scaling intrinsically prohibits its application to large time- and length-scales. Recent development of MLIPs provides new opportunities to increase computational efficiency while maintaining near DFT accuracy. The present work presents an MLIP that combines the need to include the electronic degrees of freedom with computational efficiency. 

In this work, we developed CHGNet and demonstrated the effectiveness of incorporating magnetic moments as a proxy for inferring the atomic charge in atomistic simulations, which results in the integration of electronic information and the imposition of additional charge constraints as a regularization of the MLIP. We highlight the capability of CHGNet in distinguishing Fe$^{2+}$/Fe$^{3+}$ in the study of Li$_x$FePO$_4$, which is essential for the inclusion of electronic entropy and finite temperature phase stability. In the study of LiMnO$_2$, we demonstrate CHGNet's ability to gain new insights into the relation between charge disproportionation and phase transformation in a heterovalent transition-metal oxide system from long-time charge-informed MD. 

CHGNet builds on recent advances in graph-based MLIPs \cite{dimnet_2020,Chen_Ong_2022}, but is pretrained with electronic degrees of freedom built in, which provides an ideal solution for high-throughput screening and atomistic modeling of a variety of technologically relevant oxides, including high-entropy materials \cite{Lun2020_high_entropy, sun2021_HE_catalysis}. As CHGNet is already generalized to broad chemistry during pretraining, it can also serve as a data-efficient model for high-precision simulations when augmented with fine-tuning to specific chemistries.

Despite these advances, further improvements can be achieved through several efforts. First, the use of magnetic moments for valence states inference does not strictly ensure global charge neutrality. The formal valence assignment depends on how the atomic charges are partitioned \cite{Walsh2018_chg_review}. Second, although magnetic moments are good heuristics for the atomic charge from spin-polarized calculations in ionic systems, it is recognized that the atomic charge inference for non-magnetic ions may be ambiguous and thus requires extra domain knowledge. As a result, for ions with no magnetic moment, the atom-centered magnetic moments cannot accurately reflect their atomic charges and CHGNet will infer charge from the environment similar to how other MLIP’s function. It may also be possible to enhance the model further by incorporating other approaches to charge representation, such as an electron localization function \cite{silvi1994_ELF}, electric polarization \cite{king1993_polarization_theory}, and atomic orbital-based partitioning (e.g., Wannier functions \cite{Zhang2022_wannier}). These approaches could be used for atom feature engineering in latent space.

In conclusion, CHGNet enables charge-informed atomistic simulations amenable to the study of heterovalent systems using large-scale computational simulations, expanding opportunities to study charge-transfer-coupled phenomena in computational chemistry, physics, biology, and materials science.

\section{Acknowledgments}
This work was funded by the U.S. Department of Energy, Office of Science, Office of Basic Energy Sciences, Materials Sciences and Engineering Division under Contract No. DE-AC0205CH11231 (Materials Project program KC23MP). The work was also supported by the computational resources provided by the Extreme Science and Engineering Discovery Environment (XSEDE), supported by National Science Foundation grant number ACI1053575; the National Energy Research Scientific Computing Center (NERSC), a U.S. Department of Energy Office of Science User Facility located at Lawrence Berkeley National Laboratory; and the Lawrencium Computational Cluster resource provided by the IT Division at the Lawrence Berkeley National Laboratory. The authors would also like to thank Jason Munro and Luis Barroso-Luque for helpful discussions.

\section{Methods}
\label{sec:methods}
\subsection{Data parsing}

The Materials Project Trajectory Dataset (MPtrj) was parsed from the September 2022 Materials Project database version.
We collected all the GGA and GGA$+U$ task trajectories under each material-id and followed the criteria below:
\begin{enumerate}
    \item We removed deprecated tasks and only kept tasks with the same calculation settings as the primary task, from which the material could be searched on the Materials Project website. To verify if the calculation settings were equal, we confirmed the following:
    (1) The $+U$ setting must be the same as the primary task.
    (2) The energy of the final frame cannot differ by more than 20 meV/atom from the primary task.
    \item Structures without energy and forces or electronic step convergence were removed.
    \item Structures with energy higher than 1 eV/atom or lower than 10 meV/atom relative to the relaxed structure from Materials Project's \texttt{ThermoDoc} were filtered out to eliminate large energy differences caused by variations in VASP settings, etc.
    \item Duplicate structures were removed to maintain a healthy data distribution. This removal was achieved using a \texttt{pymatgen} \texttt{StructureMatcher} and an energy matcher to tell the difference between structures. The screening criteria of the structure and energy matchers became more stringent as more structures under the same materials-id were added to the MPtrj dataset.
\end{enumerate}

Training, validation, and test sets of the MPtrj dataset were randomly selected based on the 145,923 compounds (based on the mp-id). As a result, different DFT tasks and their trajectory frames can only be included in the same set conditioned on the compound.

\subsection{Model design}
In constructing the crystal graph, the default $r_\text{cut}$ is set to 5\AA, which has been shown adequate enough for capturing long-range interactions \cite{Chen_Ong_2022}. The bond graph is constructed with a cutoff of 3\AA\ for computational efficiency. The bond distances $r_{ij}$ were expanded to $\Tilde{e}_{ij}$ by a trainable smooth radial Bessel function (SmoothRBF), as proposed in \citet{dimnet_2020}. The SmoothRBF forces the radial Bessel function and its derivative to approach zero at the graph cutoff radius, thus guaranteeing a smooth potential energy surface. The angles $\theta_{ijk}$ were expanded by Fourier basis functions to create $\Tilde{a}_{ijk}$ with trainable frequency. The atomic numbers $Z_i$, $\Tilde{e}_{ij}$, and $\Tilde{a}_{ijk}$ were then embedded into node $v^{0}_{i}$, edge $e^{0}_{ij}$, and angle features $a^{0}_{ijk}$ (all have 64 feature dimensions by default): 
\begin{equation}
\begin{aligned}
    v_i^0 &= Z_i\boldsymbol{W_v} + \boldsymbol{b_v},\\
    e_{ij,n}^0 &= \Tilde{e}_{ij}\boldsymbol{W_e}, ~ \Tilde{e}_{ij} = \sqrt{\frac{2}{5}}\frac{\sin(\frac{n \pi r_{ij}}{5})}{r_{ij}} \odot u(r_{ij})\\
    a_{ijk,\ell}^0 &= 
    \begin{split}
    \begin{cases} 
        \frac{1}{\sqrt{2\pi}} & \text{if } \ell=0 \\
        \frac{1}{\sqrt{\pi}} \cos\left[\ell\theta_{ijk}\right] &\text{if } \ell = [1,N]\\
        \frac{1}{\sqrt{\pi}} \sin\left[(\ell-N)\theta_{ijk}\right] &\text{if } \ell = [N+1,2N]
    \end{cases} , \\
    \end{split}
    \label{eq:embedding}
\end{aligned}
\end{equation}
where $\{\boldsymbol{W},\boldsymbol{b}\}$ are the trainable weights and bias. The angle is computed using $\theta_{ijk} = \arccos{\frac{e_{ij}\cdot e_{jk}}{|e_{ij}||e_{jk}|}}$. The $u(r_{ij})$ is a polynomial envelope function to enforce the value, the first and second derivative of $\Tilde{e}_{ij}$, smoothly toward 0 at the graph cutoff radius \cite{dimnet_2020}. The subscripts $n,\ell$ are the expansion orders, and we set the maximum orders for both $n$ and $\ell$ to be $2N+1=9$. The superscript denotes the index of the interaction block. The $\odot$ represents the element-wise multiplication. The edge vectors $e^t_{ij}$ are bi-directional, which is essential for $e^t_{ij}$ and $e^t_{ji}$ to be represented as a single node in the bond graph \cite{Choudhary_DeCost_2021}.

Different from previous GNN models such as M3GNet \cite{Chen_Ong_2022}, where three-body spherical harmonics features are used to update bond features, we explicitly encode and update atoms, bonds, and angles embeddings through the pair-wise connections pre-defined in atom graph and bond graph. For the atom graph convolution, a weighted message passing layer is applied to the concatenated feature vectors $(v_i^{t}||v_j^{t}||e_{ij}^{t})$ from two atoms and one bond. For the bond graph convolution, the weighted message passing layer is applied to the concatenated feature vectors $(e_{ij}^{t}||e_{jk}^{t}||a_{ijk}^{t}||v_j^{t+1})$ from two bonds, the angle between them, and the atom where the angle is located. For the angle update function, we used the same construction for the bond graph message vector but without the weighted aggregation step. The mathematical form of the atom, bond, and angle updates are formulated below:

\begin{equation}
\begin{aligned}
    v_i^{t+1} &= v_i^{t} + L_v^t \left[ \sum_{j} \Tilde{e}_{ij} \cdot \phi^t_v\left(v_i^{t}||v_j^{t}||e_{ij}^{t}\right) \right] ,\\
    e_{jk}^{t+1} &= e_{jk}^{t} + L_v^t \left[ \sum_{i} \Tilde{e}_{ij} \cdot \Tilde{e}_{jk} \cdot \phi^t_e\left(e_{ij}^{t}||e_{jk}^{t}||a_{ijk}^{t}||v_j^{t+1}\right) \right],\\
    a_{ijk,f}^{t+1} &= a_{ijk}^{t} + \phi^t_a\left(e_{ij}^{t+1}||e_{jk}^{t+1}||a_{ijk}^{t}||v_j^{t+1}\right).\\
\end{aligned}
\end{equation}
The $L$ is a linear layer and $\phi$ is the gated multilayer perceptron (gatedMLP) \cite{Xie_Grossman_2018}:
\begin{equation}
\begin{aligned}
    L(x) &= x\boldsymbol{W}+\boldsymbol{b}, \\
    \phi(x) &= \left(\sigma \circ L_{\text{gate}}(x)\right) \odot \left(g \circ L_{\text{core}}(x)\right),
    \end{aligned}
\end{equation}
where $\sigma$ and $g$ are the \texttt{Sigmoid} and \texttt{SiLU} activation functions, respectively. The magnetic moments are predicted by a linear projection of the atom features $v_i^3$ after three interaction blocks by
\begin{equation}
    m_i =\ L_m(v^3_i).
\end{equation}
Instead of using a full interaction block, the last convolution layer only includes atom graph convolution
\begin{equation}
    v^4_i = v_i^3 + \sum_{j} \Tilde{e}_{ij} \cdot \phi^3_v\left(v_i^{3}||v_j^{3}||e_{ij}^{3}\right).
\end{equation}
The energy is calculated by a non-linear projection of the site-wise averaged feature vector over all atoms $\{v_i^4\}$. The forces and stress are calculated via auto-differentiation of the energy with respect to the atomic Cartesian coordinates and strain:
\begin{equation}
\begin{aligned}
    E_{\text{tot}} &= \sum_{i} L_3 \circ g \circ L_2 \circ g \circ L_1(v^4_i) ,\\
    \Vec{f}_i &= - \frac{\partial E_{\text{tot}}}{\partial \Vec{x}_i}, \\
    \boldsymbol{\sigma} &= \frac{1}{V}\frac{\partial E_{\text{tot}}}{\partial \boldsymbol{\varepsilon}}.
\end{aligned}
\end{equation}
Overall, with four atom convolution layers, the pretrained CHGNet can capture long-range interaction up to 20 \AA\ with a small computation cost.

\subsection{Model training}

The model is trained to minimize the summation of Huber loss (with $\delta = 0.1$) of energy, force, stress, and magmoms:
\begin{equation}
\begin{aligned}
\mathcal{L}(x,\hat{x}) &= 
\begin{split}
\begin{cases} 
    0.5\cdot(x-\hat{x})^2 & \text{if } |x-\hat{x}| < \delta \\
    \delta\cdot(|x-\hat{x}| - 0.5\delta) &\text{otherwise} 
\end{cases}
\end{split}.
\end{aligned}
\end{equation}

The loss function is a weighted sum of the contributions from energy, forces, stress, and magmoms:
\begin{equation}
    \mathcal{L} = \mathcal{L}(E, \hat{E})+w_f\mathcal{L}(\boldsymbol{f}, \hat{\boldsymbol{f}})+w_{\sigma}\mathcal{L}(\boldsymbol{\sigma}, \hat{\boldsymbol{\sigma}})+ w_m\mathcal{L}(m, \hat{m}),
\end{equation}
where the weights for the forces, stress, and magmoms are set to $w_f = 1$, $w_{\sigma} = 0.1$, and $w_m = 0.1$, respectively. The DFT energies are normalized with elemental reference energies before fitting to CHGNet to decrease variances \cite{Chen_Ong_2022}. The absolute values of DFT magmoms are used for training. The batch size is set to 40 and the \texttt{Adam} optimizer is used with $10^{-3}$ as the initial learning rate. The \texttt{CosineAnnealingLR} scheduler is used to adjust the learning rate 10 times per epoch, and the final learning rate decays to $10^{-5}$ after 20 epochs.

\subsection{Software interface}
CHGNet was implemented using \texttt{pytorch 1.12.0} \cite{paszke2019pytorch}, with crystal structure processing from \texttt{pymatgen} \cite{Ong2013_pymatgen}. Molecular dynamics and structure relaxation were simulated using the interface to \texttt{Atomic Simulation Environment (ASE)} \cite{HjorthLarsen2017_ASE}. The cluster expansions were performed using the \texttt{smol} package \cite{Barroso-Luque2022_smol}.

\subsection{Structure relaxation and molecular dynamics}
All the structure relaxations were optimized by the \texttt{FIRE} optimizer over the potential energy surface provided by CHGNet \cite{Bitzek2006_FIRE}, where the atom positions, cell shape, and cell volume were simultaneously optimized to reach converged interatomic forces of 0.1 eV/\AA.

For the MD simulations of the \textit{o}-LMO to \textit{s}-LMO phase transformation, the initial structure Li$_{20}$Mn$_{40}$O$_{80}$ was generated by randomly removing Li from a Li$_{40}$Mn$_{40}$O$_{80}$ supercell of the orthorhombic structure and relaxing with DFT. The MD simulation was run under the NVT ensemble, with a time step of 2 fs at T = 1100 K for 1.5 ns. For the simulated XRD in Fig. \ref{fig:LMO}(b), the structures at 0.0, 0.3, 0.6, 0.9, 1.2, and 1.5 ns were coarse-grained to their nearest Wyckoff positions to remove noisy peaks. In Fig. \ref{fig:LMO}(c), Mn$_\text{oct}$ and Mn$_\text{tet}$ were determined by counting the number of bonding oxygen ions within 2.52 \AA. If six bonding oxygen ions were found, then the Mn ion is categorized into Mn$_\text{oct}$; if less than six bonding oxygen ions were found, the Mn ion is coarse-grained into Mn$_\text{tet}$ for representation of lower coordinated environments. In Fig. \ref{fig:LMO}(e), Mn$^{2+}$ and Mn$^{3+}$ are classified by CHGNet magmom threshold of 4.2 $\mu_B$ \cite{Barroso-Luque2022_CE_theory}.

For the MD simulations of garnet Li$_3$La$_3$Te$_2$O$_{12}$ systems, a time step of 2 fs was used. We ramped up the temperature to the targeted temperature in the NVT ensemble with at least 1 ps. Then, after equilibrating the system for 50 ps, the lithium self-diffusion coefficients were obtained by calculating the mean squared displacements of trajectories for at least 2.3 ns. The uncertainty analysis of the diffusion coefficient values was conducted following the empirical error estimation scheme proposed by \citet{He_Zhu_Epstein_Mo_2018}. In Li${_{3+\delta}}$, the excess lithium was stuffed to an intermediate octahedral ($48g$) site to face-share with the fully occupied $24d$ tetrahedral sites.

\subsection{Phase diagram calculations}
The cluster expansions (CEs) of Li$_x$FePO$_4$ were performed with pair interactions up to 11 \AA\ and triplet interactions up to 7 \AA\ based on the relaxed unit cell of LiFePO$_4$. For better energy accuracy, we first fine-tuned CHGNet with the Materials Project structures in the Li-Fe-P-O chemical space, which decreased the test error from 23 meV/atom to 15 meV/atom (see Supplementary Information). We applied CHGNet to relax 456 different structures in Li$_x$FePO$_4$ ($0\leq x\leq 1$) and predict the energies and magmoms, where the 456 structures were generated via an automatic workflow including CE fitting, canonical CE Monte Carlo for searching the ground state at varied Li$^+$ composition and CHGNet relaxation. The charge-decorated CE is defined on coupled sublattices over Li$^+$/vacancy and Fe$^{2+}$/Fe$^{3+}$ sites, where Fe$^{2+}$ and Fe$^{3+}$ are treated as different species. In addition, the non-charge-decorated CE 
is defined only on Li$^+$/vacancy sites. In the charge-decorated CE, Fe$^{2+}$/Fe$^{3+}$ is classified with magmom in $[3\mu_B,4\mu_B]$ and $[4\mu_B,5\mu_B]$, respectively \cite{Barroso-Luque2022_CE_theory}.

The semigrand canonical Monte Carlo simulations were implemented using the Metropolis--Hastings algorithm, where 20\% of the MC steps were implemented canonically (swapping Li$^+$/vacancy or Fe$^{2+}$/Fe$^{3+}$) and 80\% of the MC steps were implemented grand-canonically using the table-exchange method \cite{Xie2023_chgMC,Deng2020_TableExchange}. The simulations were implemented on a $8\times 6\times 4$ of the unit cell of LiFePO$_4$.  In each MC simulation, we scanned the chemical potential in the $[-5.6,-4.8]$ range with a step of $0.01$ and sampled the temperatures from 0 to 1000 K. The boundary for the solid solution stable phases is determined with a difference in the Li concentration $ < 0.05$ by $\Delta\mu = 0.01$ eV.

The effects of configurational and electronic entropy can be investigated via $ S(\text{Li},e) = S'(\text{Li}) + S'(e) + I(\text{Li},e)$ as described in Ref. \cite{Zhou_Maxisch_Ceder_2006}. $S'$ represents the conditional entropy $S(X|Y)$ from X (either Li or $e$) degree of freedom given fixed Y ($e$ or Li), and $I(\text{Li},e)$ denotes the mutual information of the two degrees of freedom. $S'(e/\text{Li})$ can be acquired from a canonical MC with frozen configuration of either Li/vacancy or Fe$^{2+}$/Fe$^{3+}$ ordering. This operation is facilitated by explicitly incorporating the charge decoration degree of freedom with the cluster expansion, a necessity substantiated by the atomic charge inference derived from CHGNet.

\subsection{DFT calculations}
DFT calculations were performed with the \textit{Vienna ab initio simulation package} (VASP) using the projector-augmented wave method \cite{kresse1996VASP, kresse1999PAW}, a plane-wave basis set with an energy cutoff of 520 eV, and a reciprocal space discretization of 25 \textit{k}-points per \AA$^{-1}$. All the calculations were converged to $10^{-6}$ eV in total energy for electronic loops and 0.02 eV/Å in interatomic forces for ionic loops. We used the Perdew–Burke–Ernzerhof (PBE) generalized gradient approximation exchange-correlation functional \cite{Perdew1996_PBE} with rotationally-averaged Hubbard U correction (GGA$+U$) to compensate for the self-interaction error on transition metal atoms (3.9 eV for Mn) \cite{Wang2006}. 

\section{Availability}
\subsection{Code availability}
The source code, pretrained weights, and example notebooks of CHGNet on the MPtrj dataset are available at \href{https://github.com/CederGroupHub/chgnet}{https://github.com/CederGroupHub/chgnet}.

\subsection{Data availability}
The dataset will be released after review.

\section{Competing interests}
The authors declare no competing interests.

\bibliography{references}

\end{document}